# Rapid-cycling *Re*BCO dipole magnet concept for muon acceleration

Henryk Piekarz, Simon Otten, Anna Kario, and Herman ten Kate

*Abstract*—We present a concept of the rapid-cycling 1.7 T *Re*BCO magnet for 1 to 10 kT/s ramping range as required for muon acceleration in a future muon collider presently under study. This approach is based on the 6 kA CORC-like cable constructed with 12 superconducting *Re*BCO tapes of 2 mm width. Based on theoretical prediction there is a linear scaling of the *Re*BCO cable hysteresis loss with the crossing magnetic field possibly generating power loss independent of the magnetic field ramping speed. This feature makes this *Re*BCO cable suitable for the construction of rapid-cycling magnets. In this work we outline the design of a dipole magnet for 1.7 T gap magnetic field and discuss the helium coolant parameters for the minimal electric power.

## I. Introduction

THE short muon lifetime of a few µs requires a rapid acceleration cycle to minimize muon decay losses. The energy gained in the accelerating radio frequency section per single beam path, and the desired exit energy of the beam determine the number of turns in the accelerator cycle. The accelerator magnetic field rules the length of the accelerator ring to reach the desired energy, and the accelerator magnetic field ramping speed sets the acceleration time. The acceleration cycle period must be as close as possible to the total muon circulation time to minimize the muon beam decay losses. As the muon lifetime is the shortest at the lowest energy the acceleration is arranged in stages starting with the shortest rings but of the highest magnetic field ramp rate.

## II. Design of the *Re*BCO rapid-cycling magnet

The conceptual design of the rapid-cycling *Re*BCO magnet with 2 x 3 conductor turns with 12 kA is shown in Fig. 1. It is based on the *Re*BCO magnet that was operated successfully at a ramp rate of 300 Ts [1]. The H-shaped ferromagnetic core is placed inside the cryostat thereby eliminating complication with the installation of individual cryostats around the *Re*BCO turns, but it requires a magnet core cooling water system inside the cryostat.

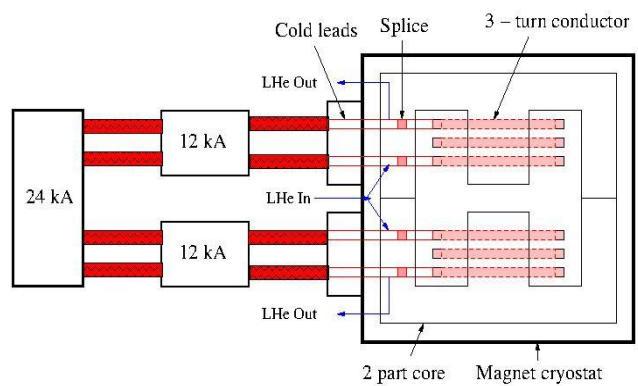

**Fig. 1** Conceptual design of the *Re*BCO magnet with 2 coils of 3 conductor turn of 2 *Re*BCO cables in parallel around an H-core carrying 12 kA per turn, 6 kA per cable, in total 72 kA cable-turns per magnet.

This arrangement eliminates the vacuum beam pipes, which are a significant source of power loss due to eddy currents generated in the pipe wall exposed to the high ramp rate magnetic field in the gap. The magnet core is constructed with 100 µm thick laminations of $Fe_3\%Si$ to facilitate the high ramp rate operations. The core saturates at 2 T, so the linear response of the magnetic field to the energizing current is up to 1.7 T. For the 30 mm beam gap as proposed for the muon accelerator dipole magnet, the energizing current for the core saturation is at 72 kA-turns. The magnet core design for the 30 mm (v) and 150 mm (h) beam gap is shown in Fig. 2. The core size of 0.45 m (h) x 0.37 m (v) requires a cryostat of 0.5 m (h) x 0.45 m (v). The horizontally split magnet core facilitates the independent installation of the coil cables on each half-core.

The *Re*BCO cable-turns occupying a narrow vertical space facilitates placement at a position with minimal exposure by the magnetic field. This also helps to install cable radiation shields to intercept the muon decays and the beam losses.

The 36-kA cable-turns per half-core can be arranged as 18 kA or 12 kA with 2-turn or 3-turn windings, respectively. For the baseline magnet design we chose the 12-kA conductor due to easier availability of the power supply and its lower cost. Each turn of 12 kA then comprises 2 *Re*BCO cables of 6 kA, which are electrically and hydraulically connected in parallel.

The warm copper current leads are attached to the cold *Re*BCO cable's core steel pipe with a bi-metallic SS-Cu pipe. The *Re*BCO cable's core steel pipe is welded to the SS section

Corresponding author: Henryk Piekarz, hpiekarz@fnal.gov, Fermi National Accelerator Laboratory, Batavia, Il 60510, USA. S. Otten (s.j.otten@utwente.nl), A.U. Kario (a.u.kario@utwente.nl; anna.kario@cern.ch) and H. H. J. ten Kate,(h.h.j.tenkate@utwente.nl; herman.tenkate@cern.ch) are with the University of Twente, 7522 NB Enschede, The Netherlands, and CERN, CH-1211, POB Geneva 23, Switzerland.
This work was funded in part by the Fermi Forward Discovery Group under Contract No. 89243024CSC000002 with the US Department of Energy (DOE), Office of Science, Office of High Energy Physic and by the European Union (E.U.), European Research Executive Agency (REA). Views and opinions are those of the authors only and do not necessarily reflect those of DOE (U.S.A.) nor REAA (E.U.).

and the copper lead end is soldered to the copper section. The *Re*BCO tapes are spliced to the lead copper pipe, which also returns helium to the cryogenic plant.

The design of the *Re*BCO CORC-like cable [2, 3] is shown in Fig.3. It consists of 3 turns of 2 cables in parallel arranged in a narrow vertical assembly. The cable is constructed with twelve 2 mm wide, 100 µm thick *Re*BCO tapes [4] wound helically in a single layer with a 100 mm pitch over the non-magnetic (316LN) supporting cooling pipe of 8 mm OD and with 0.5 mm thick wall. This cable arrangement suppresses the eddy currents induced power loss in the non-superconducting components of the *Re*BCO tape. There are 24 cable sections per magnet. Single-phase helium provides conduction cooling of the *Re*BCO tapes through the wall of the cable's cooling pipe.

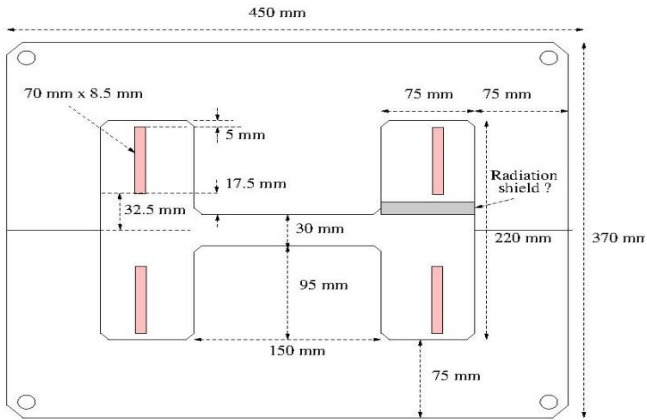

**Fig. 2.** Magnet H-core design for 2 T saturation magnetic field with a 30 mm gap where the accelerator beam passes. The coil is split into two coil sections of three turns, and each turn comprises 2 *Re*BCO cables.

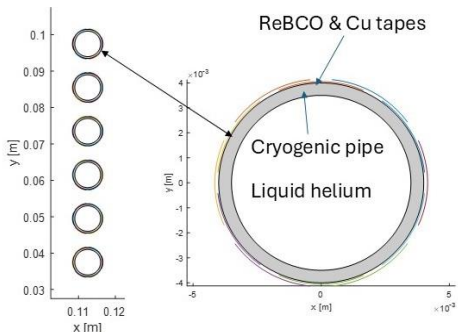

**Fig. 3.** *Re*BCO cable design for 6 kA. 12 tapes of 2 mm width wrapped around a cooling pipe of 8 mm OD in a single layer. Six cable sections are arranged in 3 turns of 2 cables in parallel per pole. 72 cable sections in total per magnet.

II. *Re*BCO CABLE AND MAGNET HYSTERESIS LOSS

In fast-ramp magnets AC losses can be significant, and performance must be studied to minimize these, and for dimensioning the necessary cooling system and power. In the case of *Re*BCO tapes, the most dominant contribution is hysteresis loss and to a much lesser extent coupling loss and eddy current loss. Hysteresis loss is proportional to the width of the *Re*BCO layer in a tape exposed to the local transverse component of the magnetic field. For this reason, the tape width applied in the cable is already 2 mm instead of the more usual 4 mm. The *Re*BCO cable must be exposed to a transverse magnetic field component to the tape surface as low as possible to minimize the hysteresis loss in magnets when used in fast-ramp accelerators. The optimized but still practical placement with minimum transverse magnetic field component is shown in Fig. 4. The maximum transverse magnetic field in the cable space averaged over the cables is 0.19 T, about 10% of the gap magnetic field. The gain in magnetic field reduction and thus in hysteresis loss by optimization of the cable positioning is significant and nearly 50%. A numerical calculation is used based on a network model [4] to calculate the hysteresis loss for each cable section in detail. For comparison, a convenient analytical expression [5] for the hysteresis loss that gives satisfactory 1st order results is:

$$Q\ hyst = B\ max\ x\ N\ x\ I_c\ x\ \frac{w}{\pi \cos(\alpha)},$$

where $N\ x\ I_c$ is the critical current per cable, σ tape lay angle, and $\frac{w}{\pi \cos(\alpha)}$, the effective tape width & length. The critical current, $I_c$ versus magnetic field data were taken from [6]. The ramping magnetic field up to 1.7 T was represented by a step function within the half-cycle of 3.4 $10^{-3}$ s, representing the magnet gap ramp rate of 1 kT/s. The magnet current waveform and the calculated hysteresis loss in the cables for one quarter magnet at 15 K operating current are shown in Fig. 5.

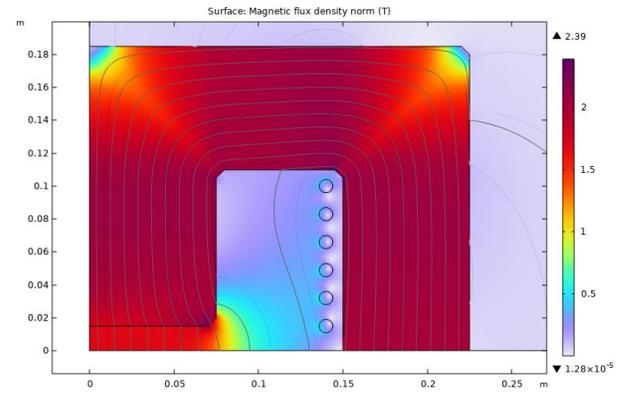

**Fig. 4.** Magnetic field calculation (in tesla) in the ferromagnetic core and the six cable sections in ¼ magnet with optimized *Re*BCO cable placement.

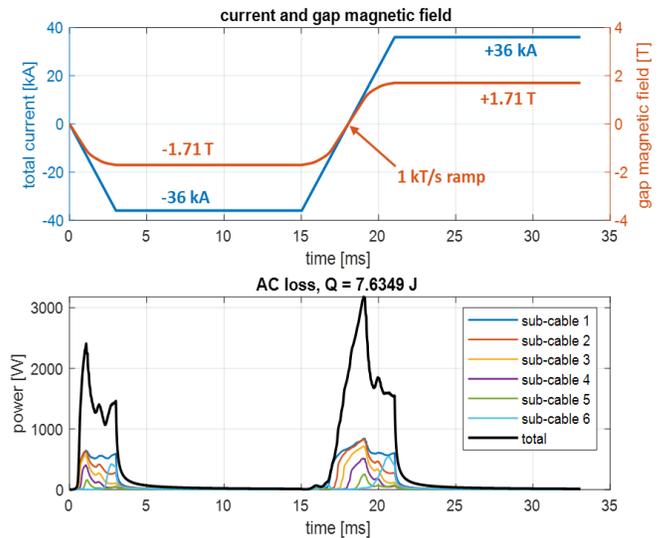

**Fig. 5.** Current in the six cables and gap magnetic field (top) and momentary hysteresis loss versus time in one quarter of the *Re*BCO magnet (bottom).

For the 12 *Re*BCO tapes per cable the 6-kA operational current constitutes 50% of the critical one. The calculated hysteresis loss of 7.8 J/m corresponds to some 61 J/m for the full cycle and full magnet. This hysteresis loss is characteristic for the cable design and its exposure to the magnetic field but in first order independent of the magnetic field ramp rate. Note that these results are illustrative for this make of the *Re*BCO tape and results may vary to some extent when other tapes are used.

### IV. CONSIDERING 5.5 K *Re*BCO OPERATION

For cooling the *Re*BCO cable it is proposed to use single-phase helium of 5.5 K at 0.28 MPa as used before [1], to take advantage of an anomalously high helium specific heat in the 5 to 6 K range [7], as shown in Fig. 6. The *Re*BCO hysteresis loss is absorbed by the flowing helium coolant, thereby increasing its temperature and enthalpy. The cryogenic loss, Q, is set by the change of enthalpy H between inlet and outlet of the conductor in combination with the helium flow rate F:

$$Q \,[J/s] = (H_{out} - H_{in}) \,[J/g] \times F \,\,\,[g/s].$$

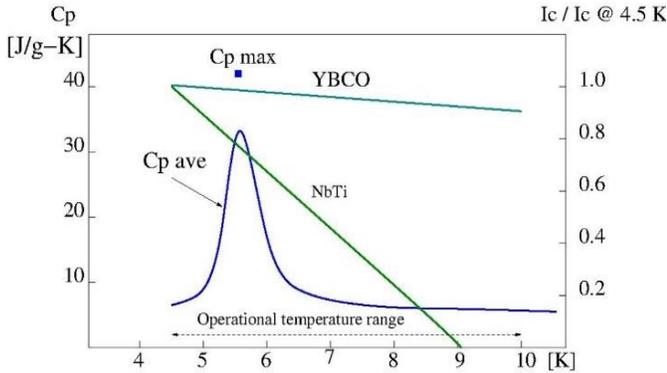

**Fig. 6.** Helium specific heat and relative critical current as function of temperature for *Re*BCO and NbTi superconductors.

This power loss must be compensated by the cryogenic plant to keep the conductor temperature constant. Enthalpy depends on helium temperature and pressure. In the coils with relatively short conductors the helium pressure is practically constant, so the change in temperature determines the change in enthalpy. The helium temperature rise induced by external heat depends on the helium specific heat, which is nearly constant in the 4 to 1500 K range. This feature allows the empirically devised formula for the needed minimum cryogenic power (Coefficient of Performance – CoP) to be based only on the difference between operational temperature and room temperature without involving the temperature dependent helium properties.

As shown in Fig. 6, however, the high helium specific heat in 5 to 6 K range strongly suppresses helium temperature rise relative to that at any other temperature range. Consequently, transfer of the heat due to hysteresis loss in the conductor to the cooling helium is minimized and leading to the reduced cryogenic power at 5.5 K relative to that at 15 K by a factor of about 6, the ratio of the corresponding specific heats.

Contrary to NbTi superconductor, the rather slow decrease in *Re*BCO critical current with rising temperature can secure stability in the 4.5 to 10 K range. The higher critical current at 5.5 K [8] relative to that at 15 K, allows to reduce the number of *Re*BCO tapes by 1/3 and requires a smaller diameter of the supporting cryogenic pipe, as shown in Fig. 7. The magnetic field descending from the core is reduced by about a factor 2 but the minimum distance to the magnet core wall is not changed, securing the same space for cable ABS holders and MLI insulation. Consequently, the hysteresis loss at 5.5 K may be reduced by about a factor of 18 (hysteresis loss x 6, magnetic field x 2, *Re*BCO tapes x 1.5) compared to that at 15 K, e.g. from some 61 J/m down to some 3 J/m per 1 kT/s full ramp.

Required electric power depends on the Coefficient of Performance (CoP) and the Cryogenic Plant efficiency. The CoP for operation at 5.5 K is 1.9% and 5% at 15 K, respectively. The cryogenic plant efficiency for the large-scale system (as needed for the muon accelerator) is about 20%. With these parameters the electric power for the operations at 5.5 K is some 4.5 kW/m and 30 kW/m at 15 K, respectively.

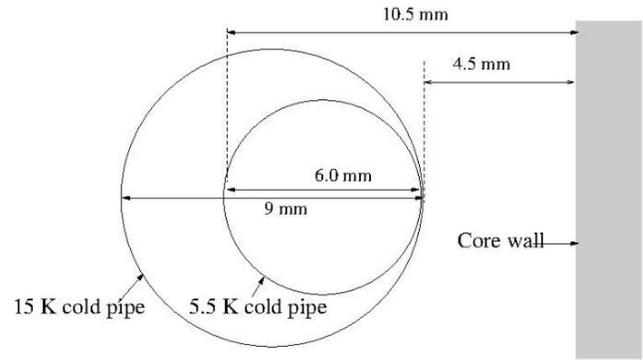

**Fig.7.** Positioning of the *Re*BCO cable cooling pipe for 15 K and 5.5 K operation.

In addition, the operation at 5.5 K may require 33% fewer *Re*BCO tapes and 33% smaller volume of liquid helium. As the *Re*BCO cable cooling efficiency depends on the helium mass flow, the helium pressure for operation at 15 K must also be increased to 1.6 MPa. It is noted that the above predictions need yet to be demonstrated in practice for which a relevant coil experiment is required.

### V. EFFICIENCY OF *Re*BCO CONDUCTOR COOLING

The high resistivity of the cooling pipe of steel supporting the *Re*BCO tapes suppresses eddy currents-based power loss, but its wall constitutes an impediment to the heat exchange between *Re*BCO tapes and the flow of helium. In periodic pulse operation the increased temperature of the *Re*BCO tapes during the short magnet excitation pulse must return to starting condition before the next current pulse comes, e.g. 0.2 s for 5 Hz operation. This is best achieved with the cable cryogenic pipe made of a high thermal diffusivity copper alloy like CuZr40, but the very low resistivity at cryogenic temperature causes a high eddy current loss as shown in Fig. 8. On the other hand, the low thermal diffusivity of the steel at cryogenic temperature significantly slows the heat transfer. Consequently, for steel or copper alloys the cable pipe wall must be as thin as possible. For the steel pipe the 0.5 mm wall is considered sturdy enough for winding the *Re*BCO tape. Based on the Barlow formula [9] the maximum allowable working pressure for a pipe of 6 mm diameter with 0.5 mm wall is 23 MPa, much exceeding the operational helium pressure of 0.28 MPa, so even a thinner wall can be considered.

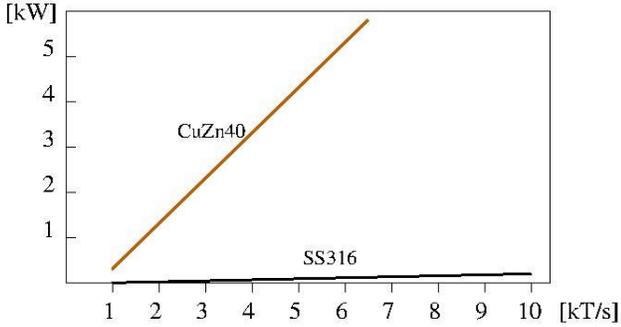

**Fig. 8.** Eddy currents loss in SS and CuZr40 cryogenic pipes per 1m magnet.

The heat diffusion time is governed by the parabolic partial differential equation [10], which projects the time to steady state of the object temperature. This time can be estimated by the formula: $t_{diff} = d^2/K$, where $d$ the distance for the heat to travel and $K$ is the temperature dependent thermal diffusion. In the temperature range of 5 to 6 K, the diffusivity of steel is $1.5 \cdot 10^{-5}$ m$^2$/s [11], so the heat transfer time across the 0.5 mm steel wall is about 0.02 s, substantially shorter than the 0.2 s period between magnetic field pulses at 5 Hz.

## VI. DISCHARGE VOLTAGE & CABLE PROTECTION

At high-ramp rate operation, the magnet conductor must withstand the repeatable high-voltage shock induced by the ramping current. For the proposed ReBCO magnet the self-inductance with the 3-turn conductor is 48 µH/m. With the 12-kA magnet energizing current the voltage shock for each half-magnet is 1700 V at 10 kT/s ramp rate. As the ReBCO cable is situated close to the magnet core wall, the cable's electrical insulation must be carefully considered. For the test magnet [1] the ReBCO cable was insulated from the magnet core wall with multiple ABS (Acrylonitrile Betadine Styrene) holders and the core itself was insulated with G10 spacers from the cryostat wall, which was electrically connected to ground. Such insulation can withstand a voltage rise much higher than 2 kV.

## VII. SUMMARY AND CONCLUSION

The independence of conductor hysteresis loss on the magnetic field ramp rate in theory suggests the possibility of conducting magnet operation at very high ramp rates. However, the projected hysteresis loss remains tentative as it is necessary to scale it with the critical current versus magnetic fields for a specific ReBCO tape conductor. The experimental verification with a short 1.7 T magnet with realistic cable and ferromagnetic core is necessary to draw conclusions for application, since in the modeling certain assumptions were made.

The dominant hysteresis loss in ReBCO cable operation needs to be minimized, in particular at temperatures other than 5.5 K. It was shown that roughly a factor 2 gain can be achieved by optimizing the position of the coil windings with respect to the ferromagnetic core thereby minimizing the lossy transverse magnetic field on the ReBCO tapes. A further option is to reduce the width of the effective ReBCO layer in the tape by applying smaller tapes (2 mm in the design presented). Next in the near future when mature, filamentary/striated tapes can be used by which another reduction of hysteresis loss by a factor of 5 to 10 may be relatively easily achieved.

An operation temperature other than the proposed 5.5 K will cause an increase in cost of both ReBCO and helium cooling.

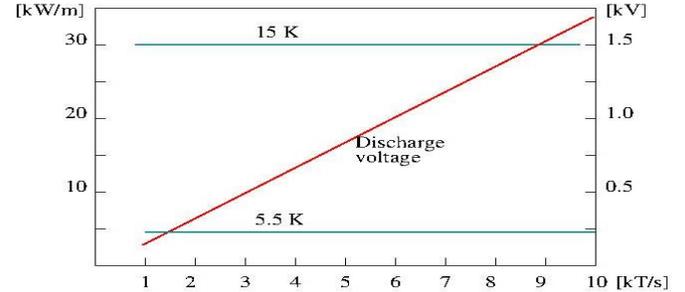

**Fig. 9.** ReBCO cable electrical cryogenic power and discharge voltage as function of magnetic field ramp rate in the 1 to 10 kT/s range for a 1 m magnet following the 3-turn cable design and operating at 5.5 or 15 K.

The necessary electric power at 5.5 K and 15 K together with the power supply discharge voltage for the 3-turn conductor design are shown in Fig. 9. Both, cryogenic power and power supply discharge voltage suggest the feasibility of the ReBCO conductor application for operation at 10 kT/s level.

For conductors in resistive magnets operating at high ramping rates, significant eddy current loss is generated in addition to the resistive loss. Copper conductors of fast cycling magnets are usually cooled with water, circulating through multiple narrow channels, thereby increasing conductor size and exposure to magnetic field variation. As a result, often the required electric power exceeds by an order of magnitude the actual conductor heating [12, 13]. As the proposed muon accelerator is planned for a low-rate power cycle, air-cooled resistive magnets with low magnetic field exposure of the single-turn conductor are considered [14, 15]. But achieving ramping rates well above 1 kT/s may not be so practical as the induced eddy current heating, in addition to resistive loss, requires high power consuming water-cooling system.

Clearly, representative demonstration magnets with both normal conducting and superconducting winding must be constructed and thoroughly tested considering all loss components and cooling power requirements in order to finally decide which technology is most suitable for the rapid cycling acceleration dipole magnets in a future muon collider.